\documentclass[aip,amsmath,amssymb,preprint,graphicx]{revtex4-1}
\usepackage{graphicx} 
\draft 

\begin{document}


\title{Towards Non-Invasive Sediment Monitoring Using Muography: A Pilot Run at the Shanghai Outer Ring Tunnel} 


\author{Kim~Siang~Khaw}
\email{kimsiang84@sjtu.edu.cn}
\author{Siew~Yan~Hoh}
\author{Tianqi~Hu}
\author{Xingyun~Huang}
\author{Jun~Kai~Ng}
\author{Yusuke~Takeuchi}
\author{Min~Yang~Tan}
\author{Jiangtao~Wang}
\author{Yinghe~Wang}
\author{Guan~Ming~Wong}
\author{Mengjie~Wu}
\author{Ning~Yan}
\altaffiliation{Present address: Department of Physics, Brown University, Rhode Island 02912, USA}
\author{Yonghao~Zeng}
\affiliation{Tsung-Dao Lee Institute and School of Physics and Astronomy, Shanghai Jiao Tong University, Shanghai 201210, China}

\author{Min~Chen}
\author{Shunxi~Gao}
\author{Lei~Li}
\author{Yujin~Shi}
\author{Jie~Tan}
\author{Qinghua~Wang}
\author{Siping~Zeng}
\affiliation{Shanghai Geological Engineering Exploration (Group) Co., Ltd, Shanghai 200092, China}

\author{Shibin~Yao}
\affiliation{Shanghai Municipal Bureau of Planning and Natural Resources, Shanghai 200001, China}

\author{Yufu Zhang}
\affiliation{Shanghai Chengtou Highway (Group) Co., Ltd, Shanghai 200335, China}

\author{Gongliang~Chen}
\author{Houwang~Wang}
\affiliation{Shanghai Surveying and Mapping Institute, Shanghai 200063, China}

\author{Jinxin~Lin}
\author{Qing~Zhan}
\affiliation{Shanghai Institute of Natural Resources Survey and Utilization, Shanghai 200040, China}

\date{\today}

\begin{abstract}
This study demonstrates the application of cosmic-ray muography as a non-invasive method for monitoring sediment accumulation and tidal influences in the Shanghai Outer Ring Tunnel, an immersed tube tunnel located beneath the Huangpu River in Shanghai, China. A portable, dual-layer plastic scintillator detector was deployed to conduct muon flux scans along the tunnel's length and to continuously monitor muon flux, allowing for the study of tidal effects. Geant4 simulations validated the correlation between muon attenuation and overburden thickness, incorporating sediment, water, and concrete layers. Key findings include a strong anti-correlation between the measured muon flux and the water levels observed at a nearby tide gauge. The results align with geotechnical data and simulations, especially in the region of interest, confirming muography's sensitivity to sediment dynamics. This work establishes muography as a robust tool for long-term, real-time monitoring of submerged infrastructure, offering significant advantages over conventional invasive techniques. The study underscores the potential for integrating muography into civil engineering practices to enhance safety and operational resilience in tidal environments.
\end{abstract}


\maketitle 

\section{Introduction}

Urban infrastructure, such as cross-river tunnels, plays a vital role in maintaining the functionality and connectivity of modern cities. Immersed tunnels are an effective method for crossing waterways, using water to transport and position tunnel segments~\cite{Lunniss:2013it}. The first immersed tunnel was constructed in 1893, and today, over 200 such tunnels have been built worldwide~\cite{Hu2015:Challenges, Pedersen2018:Fehmarnbelt}. These structures are subject to continuous geotechnical changes due to sediment accumulation, groundwater dynamics, and tidal influences, which can compromise their structural integrity over time. However, monitoring the structural health of immersed tube tunnels remains a persistent challenge in civil engineering~\cite{Jacob2022}. Traditional geotechnical methods, including borehole drilling, sonar scanning, and multibeam echo sounders (MBES)~\cite{s24134164}, are constrained by significant limitations that compromise a comprehensive risk assessment. First, the invasive nature of borehole drilling poses a risk to the structural integrity during both the installation and data acquisition. Second, these methods cannot simultaneously deliver high spatial resolution and continuous temporal monitoring, forcing a trade-off between detail and frequency. Third, techniques like MBES often necessitate operational shutdowns or hazardous, manned inspections, which disrupt critical infrastructure (e.g., tunnel operations and maritime traffic) and impede long-term risk management. These shortcomings highlight the need for innovative approaches to ensure structural resilience without compromising functionality.

In megacities like Shanghai, where rapid development coexists with aging subterranean structures, ensuring the safety and reliability of immersed tunnels has become increasingly important. The Huangpu River, a tributary of the Yangtze River flowing into the Yangtze Estuary through Shanghai Municipality, is crucial for the city's development. Its water dynamics are affected by water from the upper Taihu basin and tidal dynamics of the Yangtze Estuary. Moreover, the sediment dynamics above cross-river tunnels are highly variable, driven by natural hydrological forces such as tidal cycles in estuarine environments. Under these conditions, minor yet cumulative overburden changes -- such as uneven sediment deposition or scouring -- can lead to differential settlements and ultimately impact tunnel stability. Addressing this monitoring gap requires innovative, passive, and non-disruptive techniques that can be deployed in urban settings without disrupting ongoing operations.

Cosmic-ray muons~\cite{Nagamine:2003sv} are highly penetrating particles that can traverse hundreds of meters of rock and soil. As they pass through materials, their attenuation depends on the integrated density along their paths. By measuring the muon flux and comparing it to the expected value for a given overburden profile, changes in overburden -- such as sediment buildup -- can be identified. Muography~\cite{Bonechi:2019ckl}, utilizing these naturally occurring cosmic-ray muons to probe material density variations, offers a promising solution to this problem. This technique enables the internal imaging of large-scale structures without the use of artificial radiation or excavation. An early demonstration of muography was conducted by Eric George in 1955 to measure ice thickness above a tunnel in Australia~\cite{George:1955bzp}.

While muography has been effectively employed in archaeological investigations~\cite{Alvarez:1970ecc, Morishima:2017ghw}, volcanic imaging~\cite{Nagamine:1995np}, mine exploration~\cite{Liu:2024pxy}, and tidal-wave and tsunami monitoring~\cite{Tanaka2021, Tanaka2022}, its potential for monitoring urban infrastructure, particularly in submerged and sediment-prone environments, remains largely untapped. Recent studies have demonstrated its applicability to railway and metro tunnels for detecting hidden shafts~\cite{Thompson2020} and voids~\cite{MAO2023168391, Han_2020}; however, few investigations have examined its use in tracking temporal changes in the overburden due to natural hydrodynamic processes.

In this study, we report a pilot run of muography application to a cross-river immersed tube tunnel: the Shanghai Outer Ring Tunnel. This tunnel is a crucial segment of the Shanghai S20 expressway, which runs beneath the Huangpu River. We deployed a portable, field-ready muon detection system in two complementary modes:
\begin{itemize}
    \item a spatial scan along the tunnel axis to map variations in overburden thickness, and
    \item a fixed-point continuous monitoring system to examine the relationship between muon flux and tidal effects. 
\end{itemize}
Supporting Geant4 simulations were performed to verify the system's sensitivity to changes in overburden. Our findings indicate that muography can detect both spatial and temporal variations in overburden within a live tunnel environment, confirming its potential as a non-invasive tool for infrastructure health monitoring in dynamic urban settings. This pilot study lays the groundwork for the future deployment of muon detectors as part of smart urban sensing networks, aiming at resilient, data-driven infrastructure management.

This article is systematically structured as follows: Section~\ref{sec:outer_ring_tunnel} offers an introduction to the Shanghai Outer Ring Tunnel. Section~\ref{sec:detector_and_daq} elaborates on the muon detector and the data acquisition system utilized during the field trials. Section~\ref{sec:field_trials} outlines the measurements conducted during these trials. An analysis of the data and interpretation of the results are addressed in Section~\ref {sec:data_analysis}. Furthermore, a discussion regarding the potential applications of muography in cross-river tunnels is provided in Section~\ref {sec:discussion}, culminating in Section~\ref {sec:conclusion}, which concludes the study.

\section{Shanghai Outer Ring Tunnel}\label{sec:outer_ring_tunnel}

\begin{figure}[htbp]
    \centering
    \includegraphics[width=0.95\linewidth]{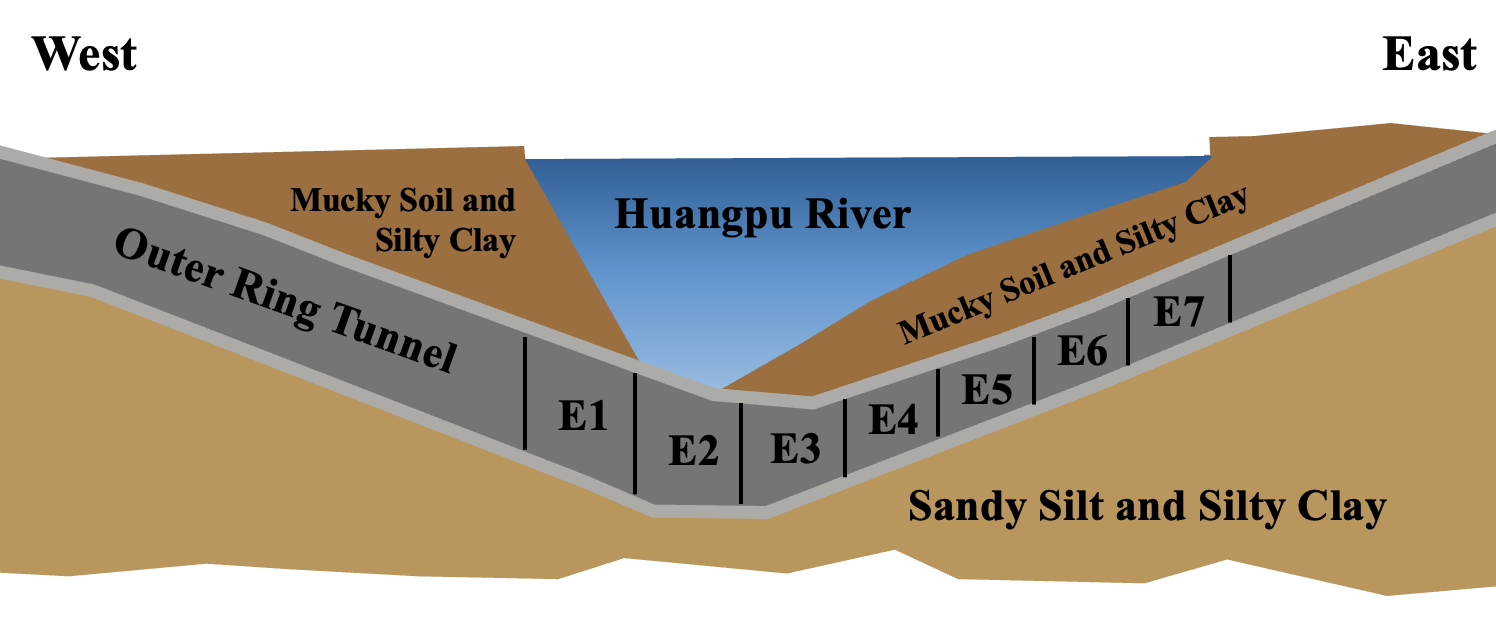}
    \caption{A simplified geological profile of the outer ring tunnel at Huangpu river~\cite{Wang2019}.}\label{fig:Outerringtunnel}
\end{figure}

\begin{figure}[htbp]
    \centering
    \includegraphics[width=0.95\linewidth]{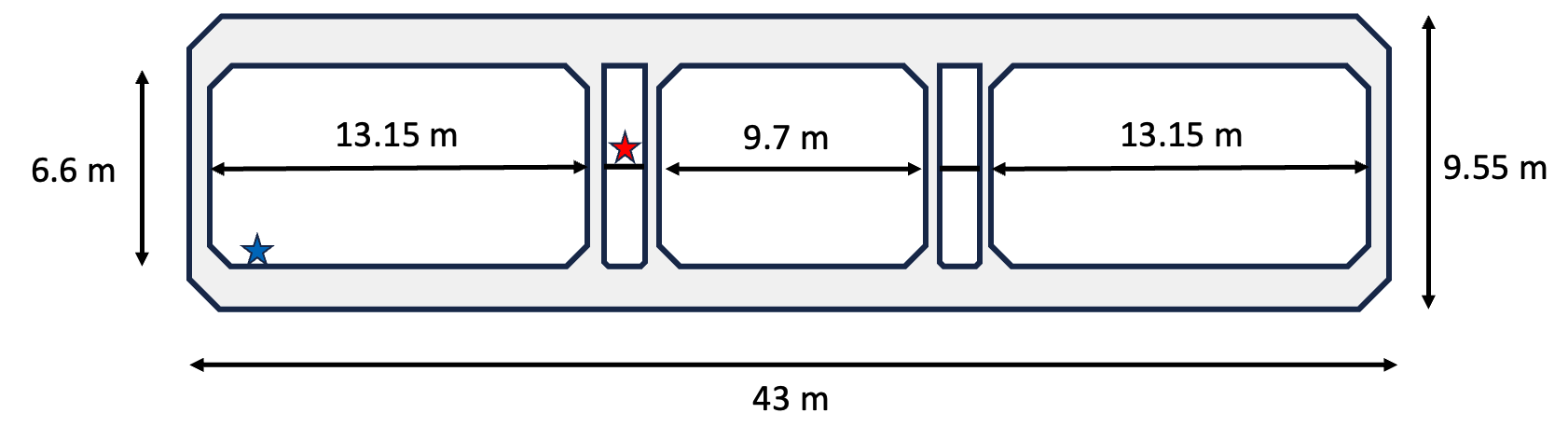}
    \caption{Transverse cross-section view of the Shanghai Outer Ring Tunnel. The blue star indicates the detector location for spatial tunnel scan, and the red star indicates the detector installation location (left service gallery) for continuous muon flux monitoring.} \label{fig:tunnel_dimension}
\end{figure}

The Shanghai Outer Ring Tunnel, as depicted in Fig.~\ref{fig:Outerringtunnel}, is an essential part of the Shanghai S20 expressway, facilitating vehicle travel between the Pudong (east of Huangpu River) and Puxi (west of Huangpu River) districts of Shanghai under the Huangpu River. As shown in Fig.~\ref{fig:tunnel_dimension}, this tunnel features a 43\,m wide, 9.55\,m high reinforced rectangular structure. It spans 2,882\,m in length and consists of three tubes that support eight lanes, along with two service galleries between the tubes dedicated to rescue operations and cabling. The submerged section of the tunnel spans a length of 736\,m and is composed of seven segments, E1 through E7. E1 begins on the western side of the tunnel. Segment E2 is located at the slopes of the tunnel site, while segments E3 to E7 gradually rise toward the eastern side.

Figure~\ref{fig:Outerringtunnel} illustrates a simplified geotechnical and geomorphic profile along the tunnel's alignment~\cite{Zhu2003, Wang2019, Jacob2022}. The terrain at the tunnel site slopes sharply to the west and gently to the east. Beneath the river lies a deep trough reaching a depth of 23.6\,m, forming an asymmetrical V-shaped riverbed. Geologically, the tunnel is situated within soil strata beneath the riverbed, comprised of sedimentary deposits commonly found in riverine environments. These overburdens consist of naturally deposited sediments, including gray sandy silt, mucky silty clay, gray mucky silty clay, muck clay, and predominantly gray mealy sand~\cite{Wang2019}, which have formed over time through weathering, erosion, and sedimentation. These sediments primarily consist of fine-grained particles displaying density variations that are influenced by factors such as organic content, compaction, and water saturation, typically ranging from 1.5 to 2.0\,g/cm$^{3}$. These materials vary in composition and properties: gray sandy silt and gray mealy sand are granular with moderate permeability, facilitating drainage; in contrast, mucky silty clay and gray mucky silty clay contain a higher clay content, leading to lower permeability and increased cohesion. Muck clay, composed of fine-grained and organic-rich material, exhibits high plasticity and water retention, which can potentially impact tunnel stability and settlement behavior.

Since its commissioning in 2003, the tunnel has exhibited signs of differential settlement, primarily caused by river flow, particularly at segment joints in the submerged sections E4-E6, resulting in water leakage into the tunnel. Moreover, the sediment thickness at the E5-E6 sections is now several meters above the design value, causing increased stress on structural stability. Additionally, tidal influences from the Huangpu River lead to periodic changes in water levels and flow velocities, complicating the surrounding environment. These hydrodynamic conditions present significant challenges for structural health monitoring
and predictive maintenance. 

These heterogeneous overburden conditions, combined with the tunnel's well-characterized structure and depth beneath the riverbed, make it an ideal natural laboratory for testing and validating muographic imaging techniques in complex urban subsurface environments.

\section{Muon Detector Design and Data Acquisition System}\label{sec:detector_and_daq}

The muon detection system is engineered for reliable field deployment in the challenging environments of the Shanghai Outer Ring Tunnel during its maintenance period from March 2024 to March 2025. The design prioritizes robustness, ease of assembly and operation, and low power consumption -- features essential for urban deployment. To that end, plastic scintillators were selected due to their mechanical durability, affordability, and light weight, making them ideal for portable and rugged use. A dual-layer configuration with two vertically spaced scintillators is used to implement a simple coincidence trigger that suppresses random noise and background radiations while enhancing muon signal purity.

The detector consists of two HDN-S2 plastic scintillators from Gao Neng Ke Di, each measuring $80 \times 20 \times 2$\,cm$^{3}$, yielding a total active area of 1,600\,cm$^{2}$. The scintillators are spaced 10\,cm apart using a custom 3D-printed support frame, resulting in an angular acceptance of 164\,degrees along the long axis and 126\,degrees along the short axis. To maximize light collection, each scintillator is coupled with light guides matched to the NVN N4021 photomultiplier tubes~\footnote{North Night Vision Technology (Nanjing) Research Institute Co., Ltd. (NVN): http://yskjnj.com/product/detail-35.html} (PMTs). A high-voltage module is also built into the PMT, enabling the detector to be operated using a commercially available 12\,V low-voltage portable power unit. 

\begin{figure}[htbp]
    \centering
    \includegraphics[width=0.95\linewidth]{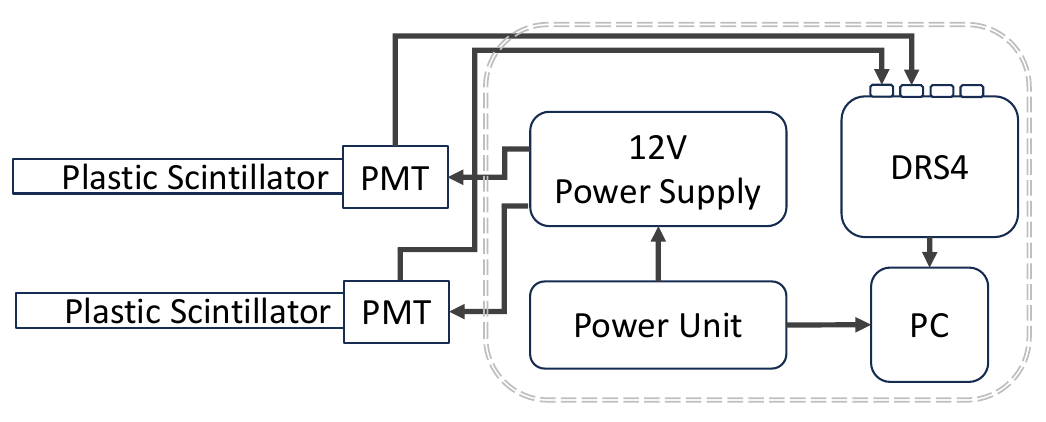}
    \caption{The schematic of the muon detection system used for tunnel spatial scan measurements. The dashed outline highlights the DAQ system used to process signals from the plastic scintillators.}
    \label{fig:setup_scan}
\end{figure}

\begin{figure}[htbp]
    \centering
    \includegraphics[width=0.95\linewidth]{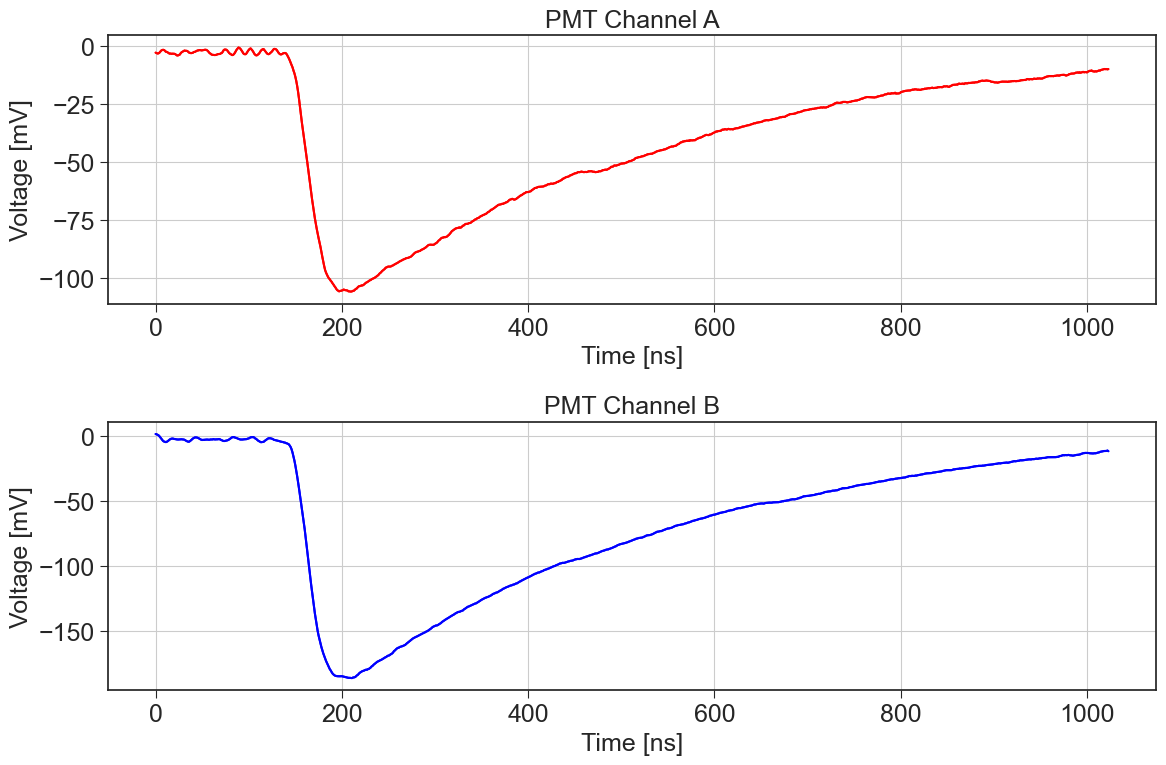}
    \caption{A typical event where both PMTs observed a double coincidence signal from the muon.}
    \label{fig:doublepulse}
\end{figure}

Two different versions of the data acquisition system (DAQ) were used during the field trials. For the spatial scan measurements used in the overburden profile study, the signals from the PMTs were digitized by the DRS4 waveform digitizer~\cite{Ritt:2010zz}, which enables data sampling at up to 5 Giga-samples per second (GSPS) with high amplitude resolution. A schematic of the DAQ system is depicted in Fig.~\ref{fig:setup_scan}. A double coincidence trigger was used to record the muon events. The threshold was set at 0.1\,V, where the most probable value (MPV) for the muon energy deposition distribution is around 0.15\,V. A typical muon event's double coincidence waveforms are shown in Fig.~\ref{fig:doublepulse}.

\begin{figure}[htbp]
    \centering
    \includegraphics[width=0.95\linewidth]{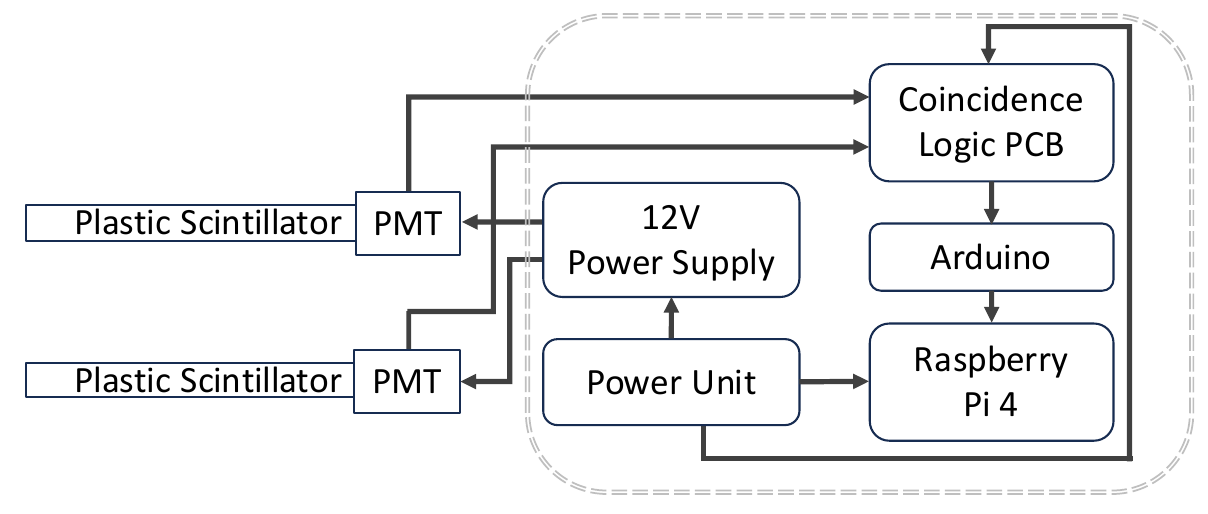}
    \caption{The schematic of the muon detection system used for the continuous tidal wave monitoring. The dashed outline highlights the DAQ system used to process signals from the plastic scintillators.}\label{fig:setup_monitor}
\end{figure}

After the spatial scan campaign, we developed a coincidence logic circuit on a printed circuit board (PCB) to replace the DRS4 for long-term muon flux monitoring. This PCB is used alongside an Arduino UNO v3 and a Raspberry Pi 4, as shown in Fig.~\ref{fig:setup_monitor}. This configuration is optimized to reduce power consumption (by replacing the laptop with a Raspberry Pi) for efficient and sustained operation. Specifically, this system removes the necessity for an on-site power source, which is often unavailable in construction or remote areas, thereby enhancing portability and suitability for long-term monitoring. Both systems are powered by a portable outdoor lithium power unit (GNV-ZC1100) with a 1\,kWh capacity, capable of delivering 1\,kW, and weighing 14.3\,kg. Prior to deployment, both detector configurations underwent calibration to ensure consistency in muon flux measurements and trigger response, enabling reliable comparison across scanning and monitoring modes. For each scintillator detector, we scanned the discriminator threshold levels while recording muon rates. The optimal operating threshold was identified at the start of the ``plateau region'' where the muon rate stabilizes despite increasing threshold voltage.

\section{Shanghai Outer Ring Tunnel Field Trial}\label{sec:field_trials}

\subsection{Muon Flux Spatial Scan in the Tunnel}

\begin{figure}[htbp]
    \centering
    \includegraphics[width=0.5\linewidth]{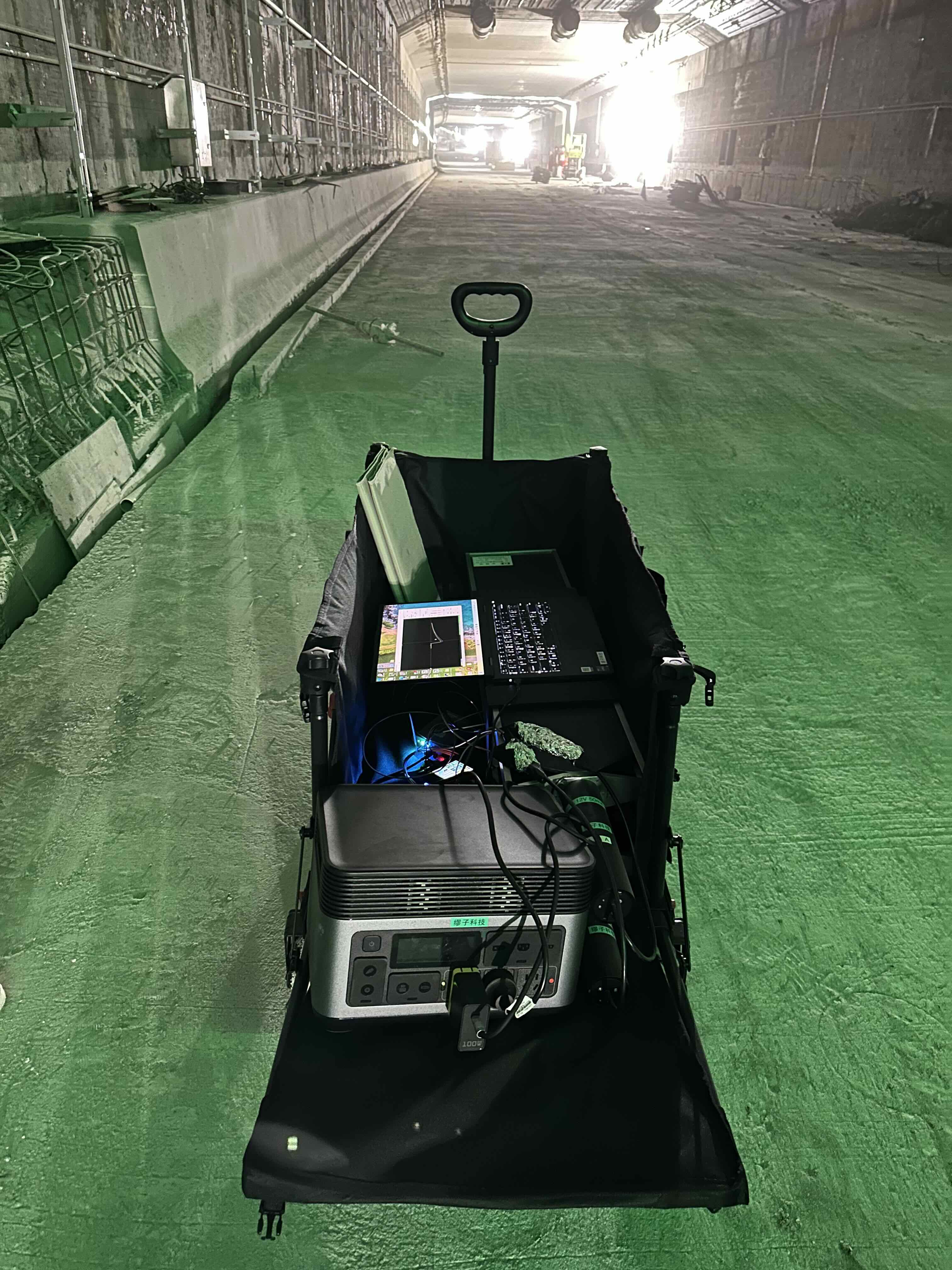}
    \caption{A photograph of the muon detection system measuring muon flux at a location during the muon flux spatial scan campaign in the Shanghai Outer Ring Tunnel.}
    \label{fig:tunnel_scan_picture}
\end{figure}

From December 26 to 27, 2024, a spatial scan of muon flux was conducted along the Shanghai Outer Ring Tunnel using the portable detection system (Fig.~\ref{fig:tunnel_scan_picture}). The operation began at segment E7 on the eastern end and proceeded westward to E1, covering the full tunnel length. Measurements were taken at 50-m intervals using a laser rangefinder, with data collected for 10 minutes at each point. The detector was consistently positioned approximately 2.3\,m from the left tunnel wall (marked by a blue star in Fig.~\ref{fig:tunnel_dimension}). Our measurement protocol requires each data acquisition run to yield a complete 10-minute cycle to qualify as valid data. Because the tunnel was under maintenance, intermittent construction activities and the passage of large vehicles occasionally disrupted data collection. In such cases, affected measurements were discarded and repeated once conditions stabilized. This guarantees that all analyzed data covers complete 10-minute intervals, maintains consistent measurement conditions for each dataset, and removes any potentially compromised data segments. The entire scan spanned about 5 hours and yielded 15 valid measurement points. We monitored the variation of muon flux at sea level during the data collection period of the tunnel scan campaign using a similar detector setup located outside the tunnel. This allowed us to apply a correction to normalize the muon count measured in the tunnel. No significant variations were observed, so no correction was applied.

\subsection{Continuous Muon Flux Monitoring}

\begin{figure}[htbp]
    \centering
    \includegraphics[width=0.95\linewidth]{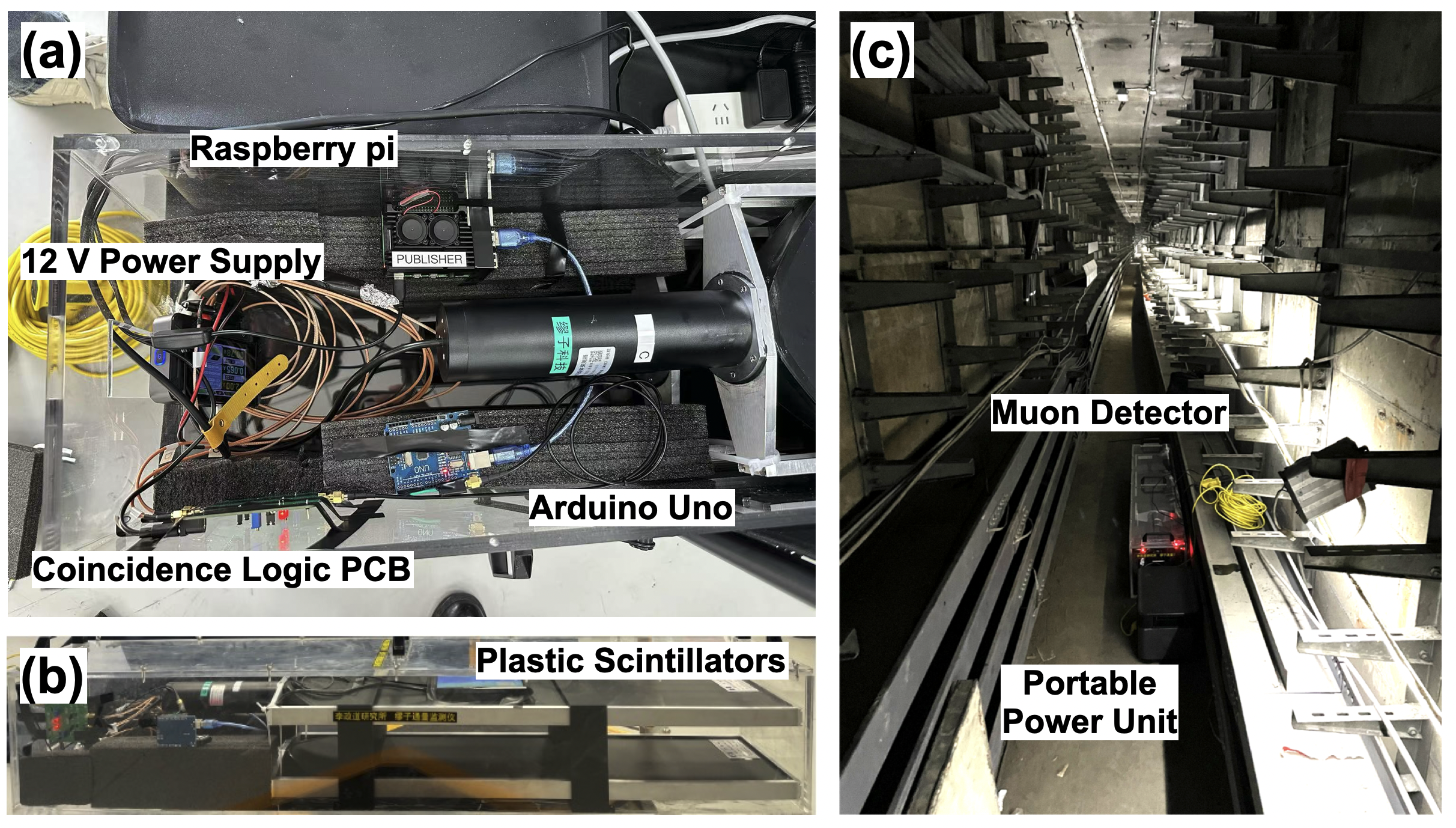}
    \caption{(a) DAQ components consist of low-cost electronics, including a Raspberry Pi, Arduino Uno, coincidence logic PCB, and power supply; (b) The dual-layered muon detector is housed in an acrylic enclosure for protection; (c) Installation of the muon detection system on the second floor of the Shanghai Outer Ring Tunnel.}\label{fig:monitor2-impression}
\end{figure}

To study the influence of tidal effects on muon flux measurement, an additional muon detection system, as shown in Fig.~\ref{fig:monitor2-impression}, was deployed to continuously monitor the muon flux at the deepest location in the tunnel, segment E2. Above the tunnel, the water column is about 20\,m. This detector was situated in the left service gallery of the tunnel (red star in Fig.~\ref{fig:tunnel_dimension}) and monitored the muon flux for 1.5\,days from January 15 to 17, 2025. This configuration significantly reduced power consumption to approximately 20 watts, allowing the muon detection system to operate continuously for over 36 hours.

\subsection{The Outer Ring Tunnel Simulation}\label{sec:ortsim}

\begin{figure}[htbp]
    \centering
    \includegraphics[width=0.9\linewidth]{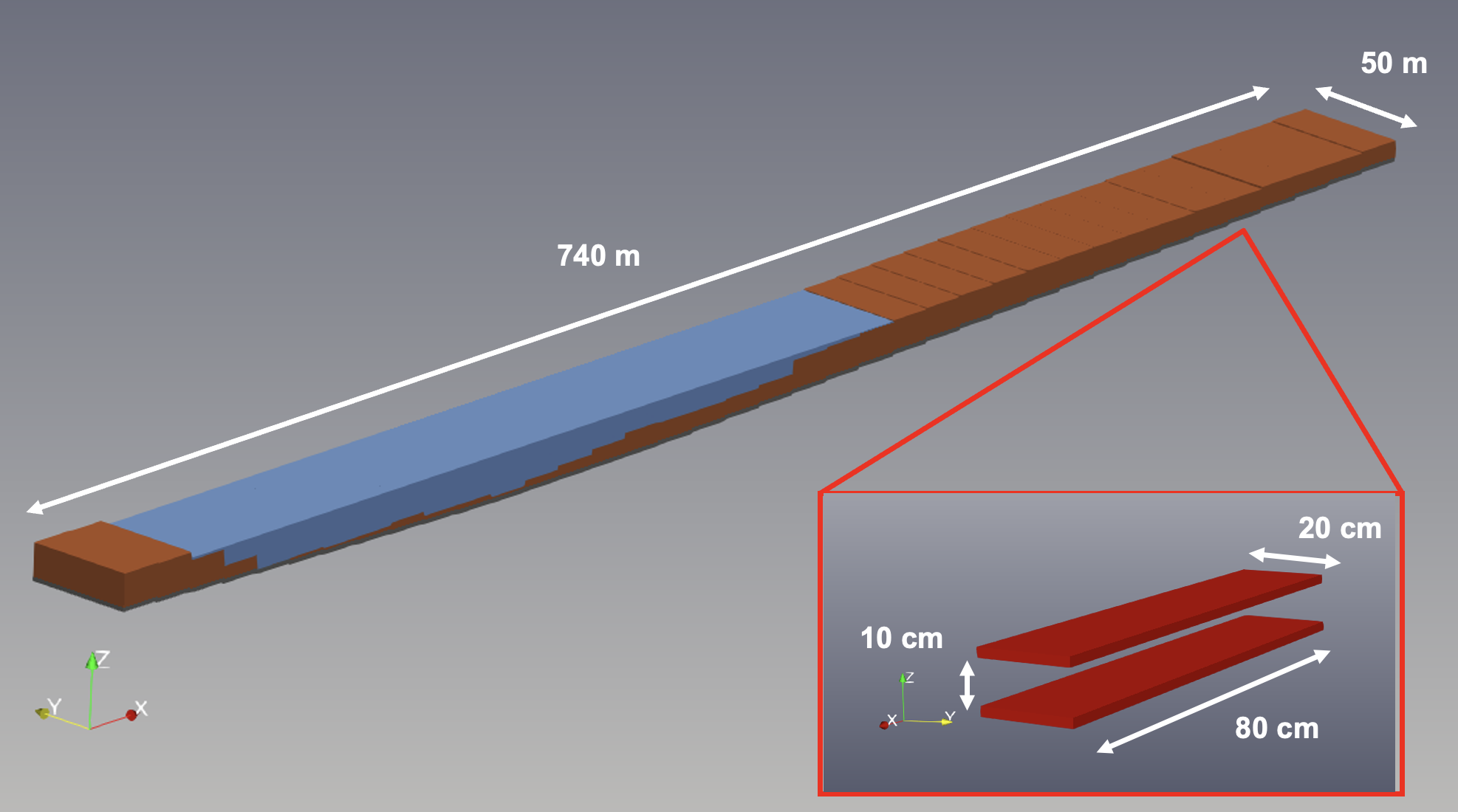}
    \caption{The simulated tunnel model, measuring 740\,m in length and 50\,m in width, consists of 37 different segmented overburden profiles made up of water (blue), sediment (brown), and concrete (black). The dual-layered detector is indicated in red box.}
    \label{fig:sim}
\end{figure}

To validate the measurement results from the scanning operation, a simplified model of the Shanghai Outer Ring Tunnel and its surroundings was constructed. The model divides the $740 \times 50$\,m$^2$ area encapsulating the tunnel into 37 equal-size discrete bins, each with an area of $20 \times 50$\,m$^2$, and varying height depending on the material budget, as shown in Fig.~\ref{fig:sim}. The material budget for the tunnel, including water, sediment, and concrete, is based on the technical drawing of the tunnel and the MBES scan conducted in 2024. The interaction of muons with the material in the model was simulated using the Geant4-based~\cite{GEANT4:2002zbu} musrSim~\cite{Sedlak:2012} package. 

\begin{figure}[htbp]
    \centering
    \includegraphics[width=0.9\linewidth]{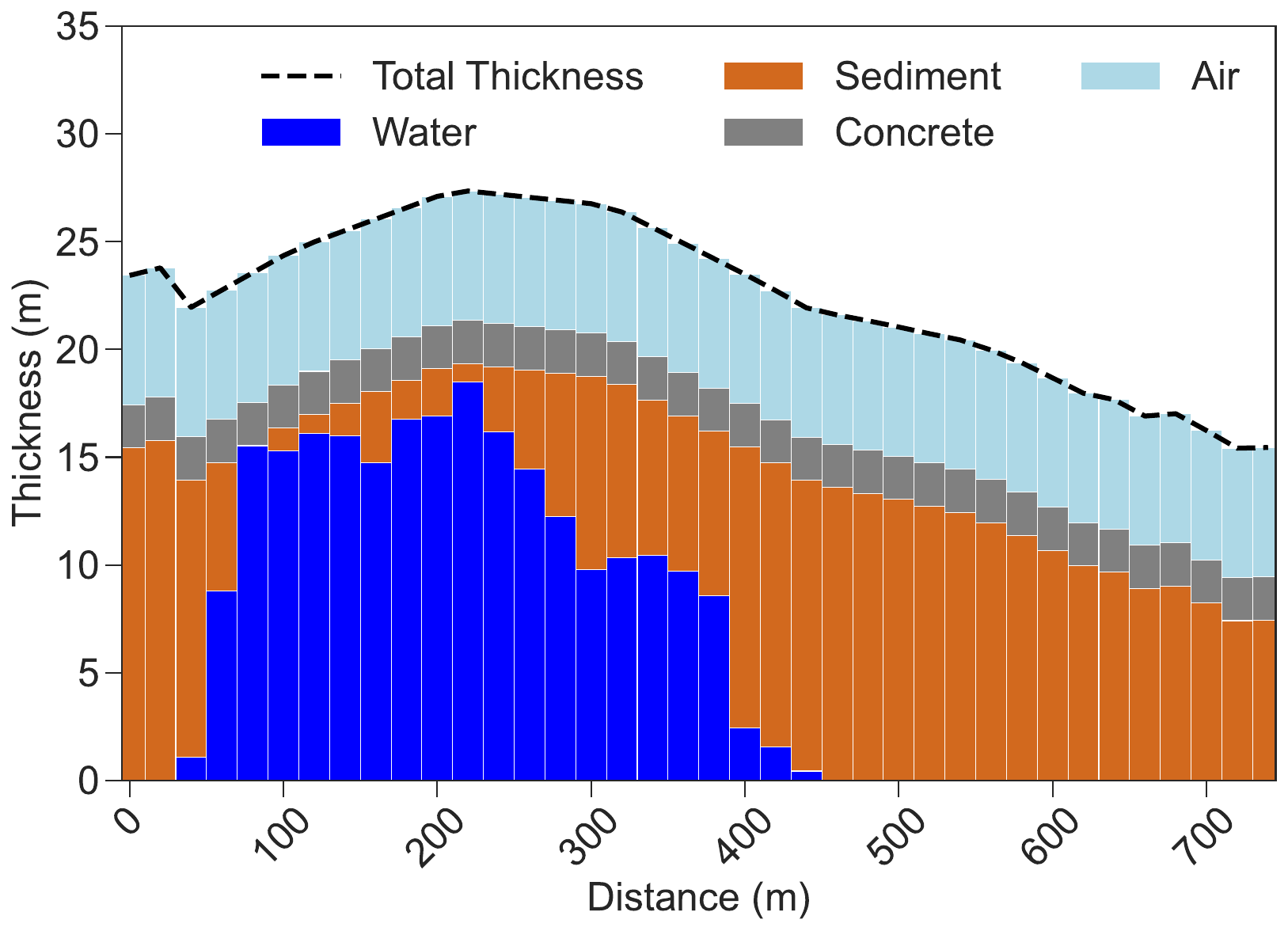}
    \caption{Material budget above the tunnel along the tunnel axis in the Geant4 simulation.}
    \label{fig:material_budget}
\end{figure}

The water density of 1.02\,g/cm$^{3}$ is selected to reflect the freshwater characteristics and typical temperature range of the Huangpu River. Water density usually fluctuates between 0.995 and 1.05\,g/cm$^{3}$, depending on temperature variations from 15 to 30\,$^{\circ}$C and local conditions. While suspended solids and dissolved particles can slightly raise water density, their overall impact remains minimal compared to other environmental factors. On the other hand, the collective sediment density of 1.8\,g/cm$^{3}$ represents the less compacted, water-saturated quality of the gray mealy sand used to simulate the sediment element. 

The third layer of the overburden profile consists of the top tunnel wall, which is made of concrete with a thickness of 2\,m and a density of 2.5\,g/cm$^{3}$. Beneath the top tunnel wall, there is an additional 6\,m of air column, representing the height of the tunnel. Subsequently, a series of muon detectors is positioned at the center of each bin to measure the expected muon flux.

Tidal effects can alter the height of the water column above the tunnel, temporarily changing the overburden thickness, thus affecting the muon flux. Although this dynamic effect is not considered in this simulation, it can be estimated using the tidal height data during the measurement period and the continuous muon flux monitoring data. In this simulation, we used the average tidal height during the measurement period. 

To simulate cosmic-ray muon events with realistic angular and energy distributions, the EcoMug generator~\cite{PAGANO2021165732} is used to generate cosmic-ray muons from a sky-plane, emitting at a rate of 129 Hz/m$^{2}$, corresponding to an exposure of about 1\,min. The attenuation of muons due to overburden above the tunnel is studied by positioning a plane source of size $800 \times 50$\,m$^{2}$ at a height of 14\,m above the tunnel. 

\section{Data Analysis and Interpretation}\label{sec:data_analysis}

\subsection{Temporal Correlation with Tidal Effects}

\begin{figure}[htbp]
    \centering
    \includegraphics[width=0.75\linewidth]{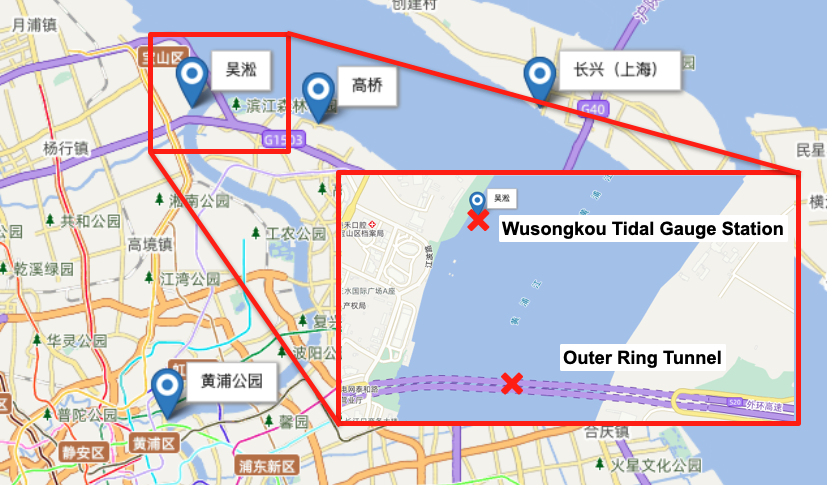}
    \caption{The location of the Wusongkou tidal gauge station (31$^{\circ}$~22$^{\prime}$~56.2$^{\prime\prime}$~N, 121$^{\circ}$~30$^{\prime}$~18.6$^{\prime\prime}$~E) and the Outer Ring Tunnel (31$^{\circ}$~22$^{\prime}$~37.6$^{\prime\prime}$~N, 121$^{\circ}$~30$^{\prime}$~23.1$^{\prime\prime}$~E).}\label{fig:tidalgeo}
\end{figure}

Prior to interpreting the spatial scan results, we first examined the temporal correlation between muon flux and tidal effects. The variation in tidal height influences the thickness of the water layer, which in turn, significantly affects the muon absorption rate by altering the overall overburden budget. To assess this relationship, tidal height data from the Wusongkou tidal gauge station (approximately 3\,km north of the Shanghai Outer Ring Tunnel; Fig.~\ref{fig:tidalgeo}) were time-matched with 10-minute muon flux measurements for a given date.

\begin{figure}[htbp]
    \centering
    \includegraphics[width=\linewidth]{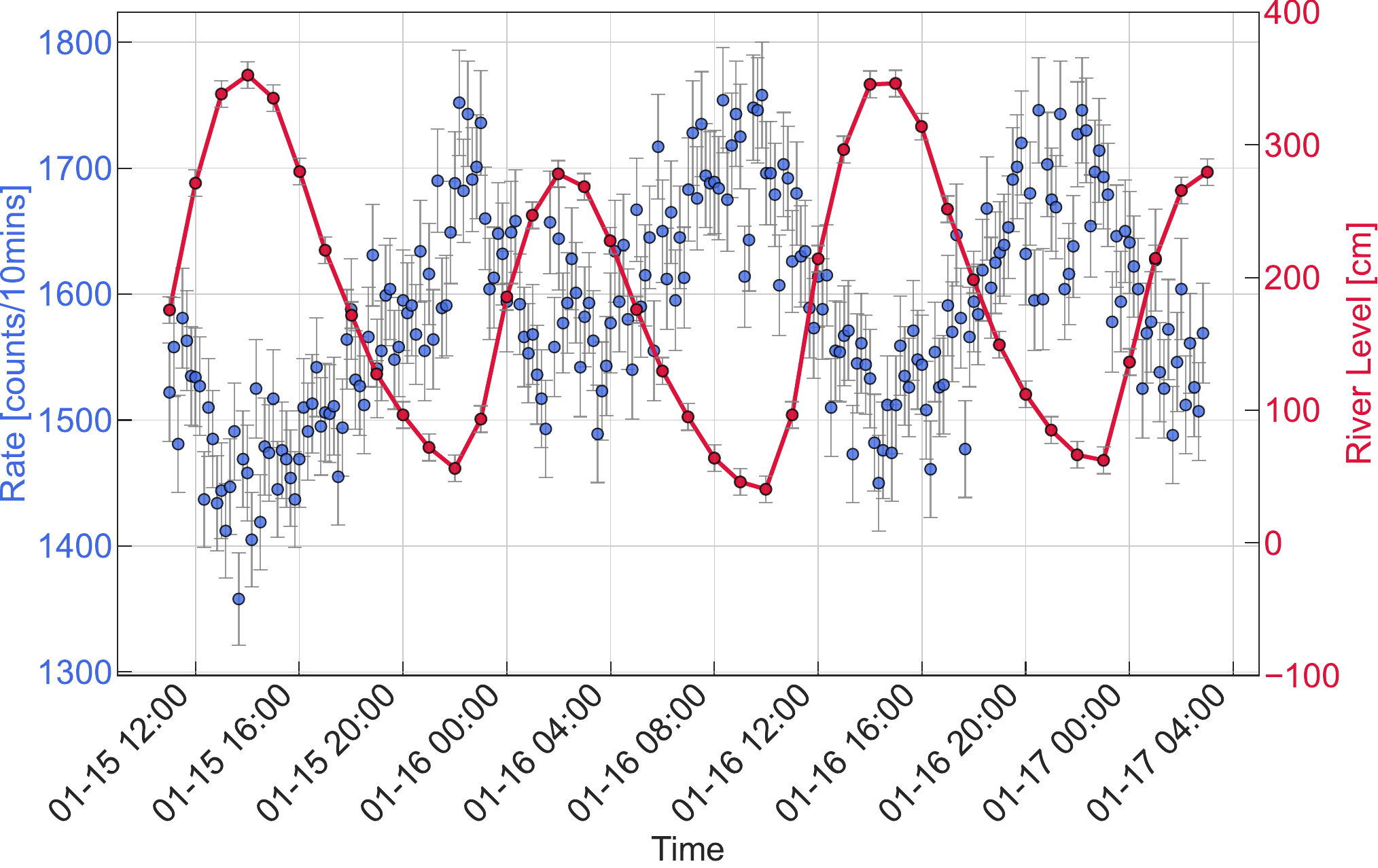}
    \caption{Time-sequential plot of the number of muon counts collected every 10 minutes (dark blue dots) and the river water level (202 cm below mean sea level) measured at the Wusongkou tidal gauge station (red curve). Error bars indicate statistical uncertainty.}
    \label{fig:tidal_muon}
\end{figure}

\begin{figure}[htbp]
    \centering
    \includegraphics[width=\linewidth]{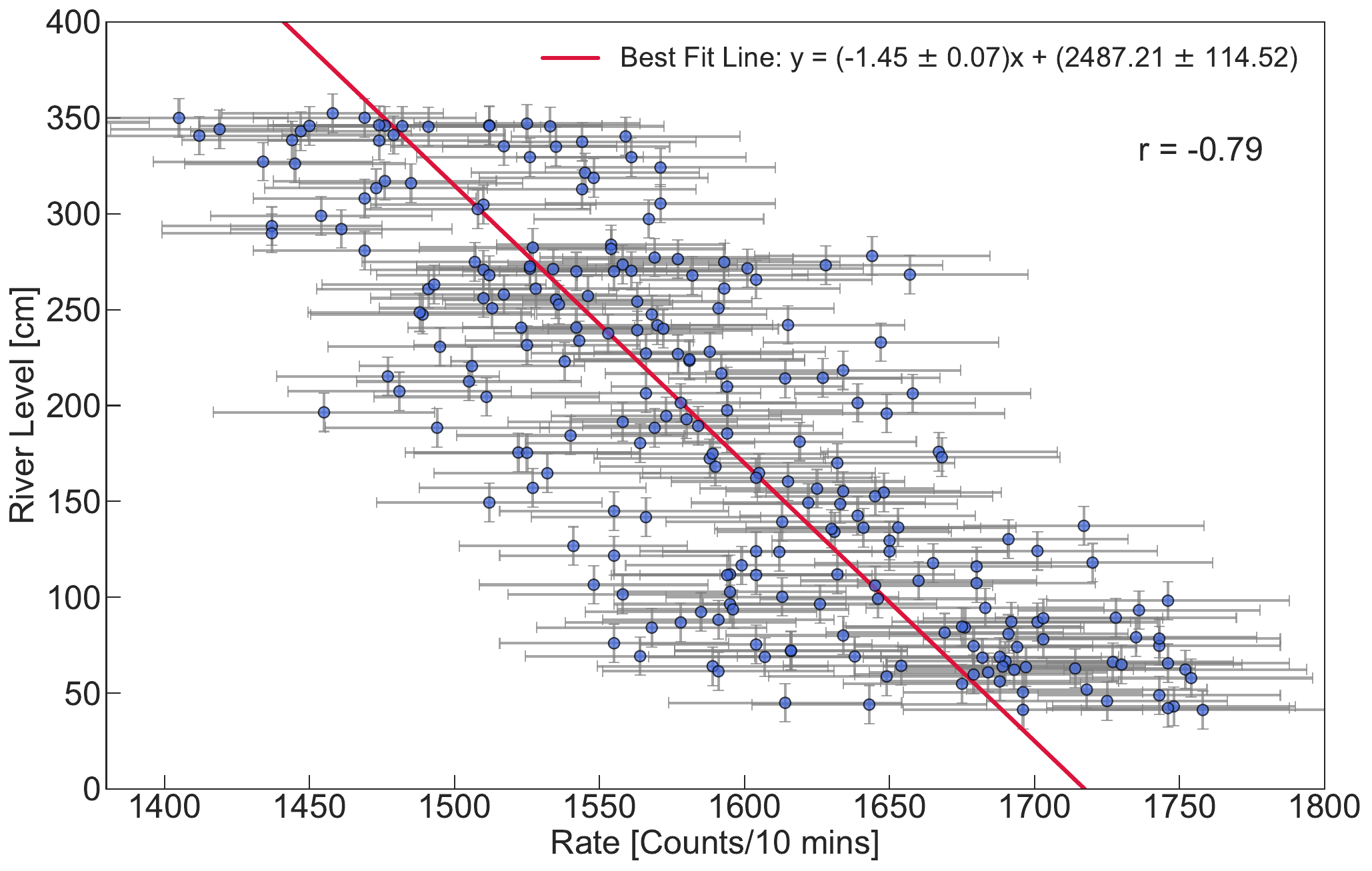}
    \caption{A scatter plot of river water levels and measured muon flux per 10 minutes. The correlation coefficient is measured at -0.8.}\label{fig:correlation}
\end{figure}

Fig.~\ref{fig:tidal_muon} shows the temporal evolution of both muon counts and astronomical tide height (ATH) over the observation period. A clear anti-correlation is observed: muon flux decreases as the tide rises and increases as it falls and vice versa. In other words, the periodic fluctuations in muon flux correspond to the tidal cycles, demonstrating an approximately 4\% decrease (or 69\,counts/10\,mins) in muon flux for each meter increase in water level, consistent with theoretical expectations. The measured correlation between muon flux and ATH revealed a strong anti-correlation relationship, as shown in Fig.~\ref{fig:correlation}. Interestingly, this periodic modulation of muon flux is analogous to the observations from the Tokyo Bay TS-HKMSDD experiment~\cite{Tanaka2022}, which reinforces the importance of accounting for tidal effects when interpreting muographic data in aquatic or sediment-rich environments.

\subsection{Spatial Variation in Muon Flux}

\begin{figure}[htbp]
    \centering
\includegraphics[width=0.95\linewidth]{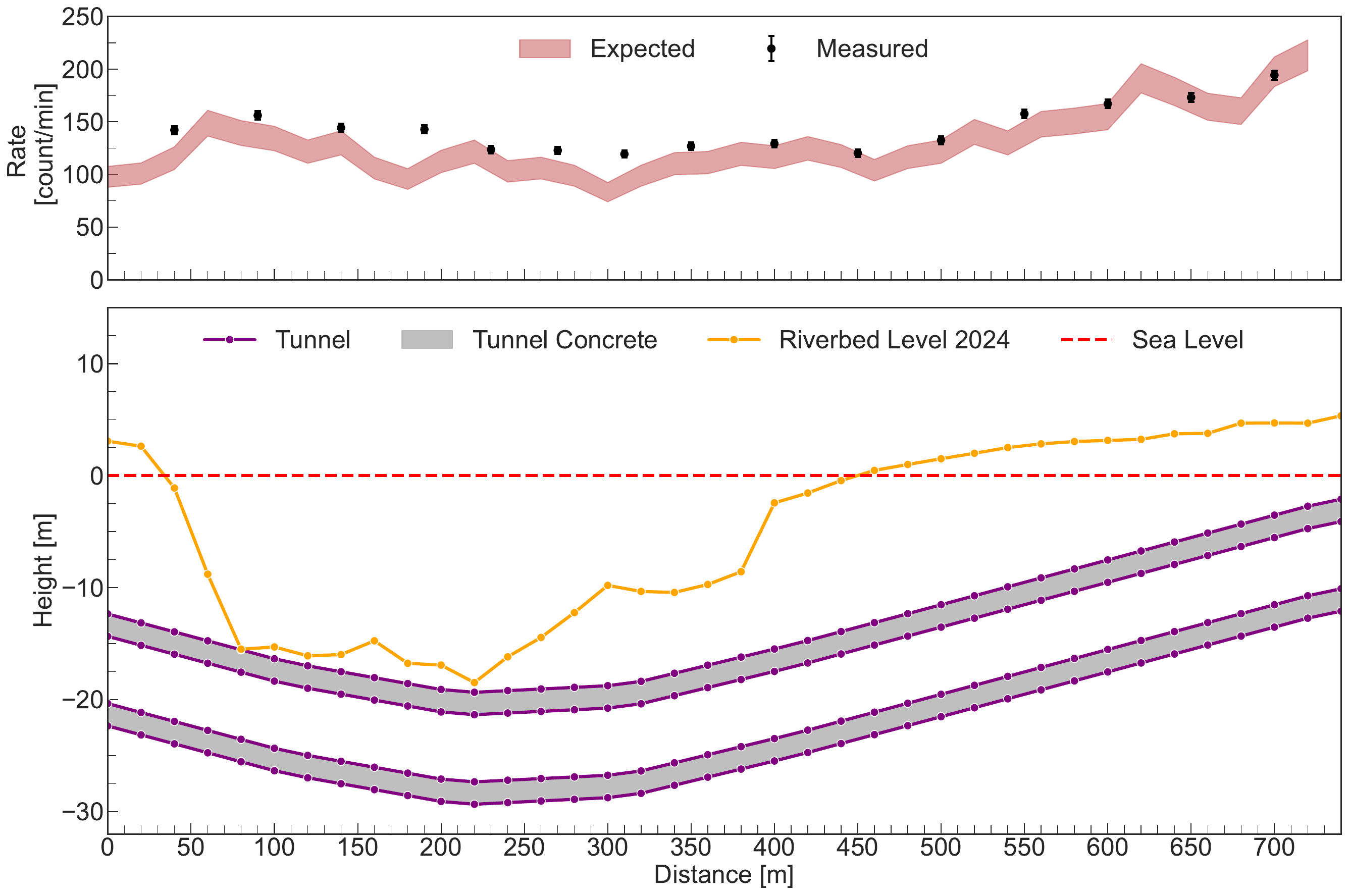}
    \caption{Top: Measured muon flux at 15 locations along the tunnel length. Bottom: Current riverbed profile measured with MBES in 2024 along the tunnel direction. The errors indicated on the muon flux rate are purely statistical.}\label{fig:muonflux_and_riverbed}
\end{figure}

The collected muon dataset from the Spatial Scan campaign was normalized to counts per minute and compared alongside the overburden profile as a function of distance, starting from the east entrance of the immersed tube tunnel at E1, as shown in Fig.~\ref{fig:muonflux_and_riverbed}. The flux systematically varies along the tunnel axis, correlating with tunnel depth and sediment thickness. 

Overall, the measured flux aligns well with the simulated flux along the tunnel axis. The variation in muon flux is less pronounced compared to the overburden profile, mainly due to the detector's large acceptance angle (allowing for a broader range to be mapped) and the relatively small integrated density along the muon path, given the changing tunnel depth. Notably, the measured and expected muon fluxes show excellent agreement in the region dominated by sediment, which is also the region of interest (E4-E6) in this study, specifically from 350\,m to the right (west). In contrast, the measured flux is consistently higher in the area dominated by river water.

The systematic differences between the expected and measured muon flux at the deepest part of the tunnel can be caused by various potential sources of error or approximation in the simulation and environmental conditions. One factor is that the measured riverbed has an uncertainty of $\pm$0.5\,m and is not shown in Fig.~\ref{fig:muonflux_and_riverbed}. If the actual riverbed at a single point is 0.5\,m lower, this means the expected muon flux should increase by about 2\%, according to the results from the tidal effect analysis. Since we used the average tidal height for the water budget in the simulation, the muon flux might be underestimated due to the decrease in tidal height during the last few measurement points. However, based on the tidal gauge data, the water level decreased by about 25\,cm per hour, which corresponds to roughly a 1.0\% increase in muon flux. This cannot fully explain the difference between the simulation and the measurement in Fig.~\ref{fig:muonflux_and_riverbed}. Another factor is the simplification of material properties in the simulation, such as assuming uniform density for both sediment and water. In reality, the densities of sediment and water can vary due to factors like organic content, compaction, and water saturation, which are not fully accounted for in the model. The next step is to develop a more accurate model of the tunnel and its surrounding environment.

\section{Discussion}\label{sec:discussion}

Building on the results of this proof-of-principle pilot study, future work will directly address the technical limitations, particularly the limited spatial resolution resulting from the detector’s wide angular acceptance and the absence of muon tracking. These constraints hampered our ability to isolate muons passing through specific regions of interest and prevented the effective use of tomographic inversion techniques.

To address these limitations, we plan to develop an improved detector prototype with finer segmentation and multiple layers to support basic muon tracking. Simulation studies are currently underway to optimize the geometry and configuration, with the goal of achieving the spatial and angular resolutions necessary for accurate path reconstruction. These enhancements will enable us to implement more advanced data reconstruction methods, including voxel-based tomographic inversion (see, for example, \cite{Tarantola:2005inverse,Menke2018:geophysical,Tassiopoulou2024:algorithms,Allner2019:metric}). Additionally, machine learning algorithms for sparse or noisy muographic data can also be incorporated.

On the analysis side, we will integrate high-resolution 3D models of the tunnel along with the geotechnical and geomorphic profiles of the Shanghai Outer Ring Tunnel~\cite{Wang2019} to improve the accuracy of sediment thickness estimates. These models will serve two main purposes: first, as input for forward simulations to improve estimates of expected muon flux; second, as a foundation for inversion workflows to map sediment thickness variations along the tunnel’s length. These improvements will enable more quantitative and spatially detailed assessments of subsurface changes. The next phase of deployment will also expand the detector network to include in-tunnel measurements and additional cross-river tunnels beneath the Huangpu River. This strategy builds on two key insights from the current study:
\begin{itemize}
\item{the observed 4\% muon flux decrease per meter of tidal rise, which confirms the sensitivity of the system to water-equivalent overburden;}
\item{the spatial variation in muon flux correlated with sediment-heavy regions, which highlights the need for improved resolution.} 
\end{itemize}

\section{Conclusion}\label{sec:conclusion}

This study presents a preliminary application of muography as a robust method for monitoring sediment thickness and assessing tidal influences within the Shanghai Outer Ring Tunnel. Through a combination of spatial scanning and fixed-site muon flux measurements, we demonstrate the method's reliability in detecting variations in overburden in a complex subsurface environment. Findings indicated a strong alignment between measured muon flux and theoretical predictions, confirming the accuracy of Geant4 simulations in replicating observed muon attenuation in the area of interest. 

The observed anti-correlation between muon flux and tidal variations highlights the strong potential of muography as a non-invasive tool for monitoring changes in subsurface infrastructure. By enabling the detection of overburden variations and sediment accumulation without physical intrusion, this technique directly supports the long-term integrity and safety assessment of critical underground structures, such as cross-river tunnels.

This study demonstrates that detector technologies developed in particle physics are already capable of meeting the resolution demands of geophysical monitoring. With continued optimization for cost-efficiency and deployment, muography could play a transformative role, complementing or even replacing conventional methods in the field of geophysical monitoring. Ultimately, this work establishes a clear path forward for integrating muography into routine infrastructure management and continuous subsurface diagnostics.

\begin{acknowledgments}

We wish to express our heartfelt gratitude to Jining Chen from the Shanghai Municipality for proposing the innovative application of muography in infrastructure studies within the Shanghai region. Our special appreciation extends to Yuxin Zhang from the Shanghai Municipal Bureau of Planning and Natural Resources and Jie Zhang from Shanghai Jiao Tong University for initiating this interdisciplinary research collaboration. We sincerely thank the staff at Shanghai Tunnel Engineering Co., Ltd., for their invaluable assistance with muon flux measurements and the deployment of the long-term monitoring detectors. We thank Samip Basnet for his helpful suggestions to improve the manuscript. We would also like to extend our gratitude to the administrative staff—Shushu Li, Rongrong Zhang, Sheng Li, Jinghua Shi, Xiaoqian Jin, Xiaojun Lu, and Mengzhu Lu—at the Tsung-Dao Lee Institute, Shanghai Jiao Tong University, for their support in establishing the TDLI Muography Group. This work was financially supported by the Tsung-Dao Lee Institute Special Research Grant for Muography Applications in Shanghai, as well as the Double First Class Start-up Fund provided by Shanghai Jiao Tong University.
\end{acknowledgments}

\bibliography{mybib}

\end{document}